\documentclass[runningheads]{llncs}
\usepackage{makeidx}
\usepackage{graphicx}
\usepackage{amsmath}
\usepackage{listings}
\usepackage{array}
\usepackage{cite}
\usepackage{algpseudocode,algorithm,algorithmicx}
\usepackage[T1]{fontenc}
\usepackage[colorinlistoftodos]{todonotes}
\usepackage{xcolor}
\usepackage{adjustbox}
\usepackage{svg}
\usepackage{rotating}
\usepackage{booktabs}
\usepackage[numbers]{natbib}

\newcommand{\vc}[1]

\usepackage{makecell}

\begin{document}

\pagestyle{headings}

\mainmatter

\title{Efficient allocation of image recognition and LLM tasks on multi-GPU system}

\author{Marcin Lawenda \inst{1} \and Krzesimir Samborski\inst{1} \and Kyrylo Khloponin\inst{1} \and Łukasz Szustak \inst{2} }
\authorrunning{M. Lawenda and K. Samborski and K. Khloponin and Ł. Szustak}
\titlerunning{Efficient allocation of image recognition and LLM tasks ...}

\institute{
Poznan Supercomputing and Networking Center, Jana Pawła II 10, 61-139 Poznań, Poland \\
\email{\{lawenda,ksamborski,kkhloponin\}@man.poznan.pl} \\
\and
Czestochowa University of Technology, Dąbrowskiego 69, 42-201 Częstochowa, Poland \\
\email{lszustak@icis.pcz.pl}
}

\maketitle

\section*{Abstract}
This work is concerned with the evaluation of the performance of parallelization of learning and tuning processes for image classification and large language models. For machine learning model in image recognition, various parallelization methods are developed based on different hardware and software scenarios: simple data parallelism, distributed data parallelism, and distributed processing. A detailed description of presented strategies is given, highlighting the challenges and benefits of their application. Furthermore, the impact of different dataset types on the tuning process of large language models is investigated. Experiments show to what extent the task type affects the iteration time in a multi-GPU environment, offering valuable insights into the optimal data utilization strategies to improve model performance. Furthermore, this study leverages the built-in parallelization mechanisms of PyTorch that can facilitate these tasks. Furthermore, performance profiling is incorporated into the study to thoroughly evaluate the impact of memory and communication operations during the training/tuning procedure. Test scenarios are developed and tested with numerous benchmarks on the NVIDIA H100 architecture showing efficiency through selected metrics.


\section{Introduction}
The rise of Machine Learning (ML) along with Large Language Models (LLMs) has revolutionized various domains such as natural language processing, computer vision, and data analysis. Known for their extensive parametrization and complex structures, these models require significant computational resources for both training and inference. Multi-card GPU systems, which leverage the parallel processing capabilities of multiple graphics processing units (GPUs), have become the essential infrastructure to meet these computational requirements \cite{lin2024universalperformancemodelingmachine}. This approach has resulted in the development of code specifically designed for parallelization, enabling faster training of much larger models by leveraging the increased computational power and memory of additional accelerators. However, efficient task allocation in multi-card GPU systems remains a significant challenge due to the inherent complexity of the model architecture, data dependencies, and hardware heterogeneity \cite{10.1145/1058129.1058148}.

The primary objective of this work is to analyse the distribution of data chunks in single and multi-GPU environments to enhance the processing efficiency of machine learning applications. Through experiments, different techniques are evaluated for scalability and execution capabilities, leading to improved planning and recommendations for future applications and enhancements. The study involves GPU benchmarking with neural models and data distribution strategies to optimize training processes. The aim is to assess the scalability in a multi-GPU setup to understand the performance of hardware-software configurations in large-scale computing. Selecting the most suitable approach for a specific application's workflow is a key challenge, especially when utilizing multiple GPU cards across nodes. The findings from these tests will offer valuable insights for refining models and tools in the future.

% Chapter 2 includes outlines the testing environment, detailing both the hardware platform and software stack utilized to enhance procedures and monitor metrics related to the time taken for specific operations within the program code. Chapter 4 is dedicated to analyzing the performance of image recognition tasks within the MobileNet2 model using the MNIST dataset, employing three distinct task distribution methods. Two optimization techniques, namely "floating operation precision" and "$pin\_memory$," were implemented. The effects of these same optimization techniques on large language models (LLMs) are discussed in Chapter 4, where the analysis considers various datasets and variations in memory and communication metrics, paralleling the approach taken for image recognition. Finally, Chapter 5 provides a summary of the research conducted, evaluating the effectiveness of the individual solutions in terms of computational and communication performance.

\section{Experimental environment and methodology}\label{section:Methodology}

\subsection{Hardware infrastructure}
The hardware platform utilized for this research is built on the HPE Cray XD665 system, which comprises of 2 AMD EPYC 9334 CPUs and 4 NVIDIA H100-94 SXM5 GPUs. There are a total of 64 CPU cores, based on Zen 3 architecture and 2.7 GHz clock speed. The system's RAM capacity per node amounts to 768 GB, complemented by NVIDIA H100-94 SXM5 GPUs, each equipped with 94GB of HBM2e memory, resulting in a total HBM2e memory capacity of 376 GB per node. The total GPU computational power across all four GPUs in each node is rated at 268 TFLOPS. The GPUs are interconnected via NVIDIA NVLink NV6.

The environment is running on Rocky Linux 9.3 with a kernel version of k08r01s02.novalocal (version 5.14.0-362.24.1.el9\_3.0.1) with CUDA 12.4, Nsight system version 2024.4.1 and SLURM version 23.11.6.

\subsection{Benchmarking software and procedures}

\subsubsection{Implementation frameworks.}

The fine-tuning of Large Language Models was conducted utilizing TorchTune - a library within the PyTorch ecosystem. Torchtune uses recipes for single-device and distributed tuning, depending on whether one or more GPUs were used. These recipes can be subsequently modified using custom configuration files in order to perform a wide array of tests\cite{TorchtuneConfigs}.

\subsubsection{Profiling tools.}
NVIDIA Nsight Systems is a comprehensive system-wide performance analysis tool designed to optimize and profile applications running on NVIDIA GPUs. It provides detailed insights into CPU and GPU interactions, memory usage, and the performance of CUDA kernels and APIs. By capturing a holistic view of the system's behavior, nsys helps identify bottlenecks, analyze workload distribution, and optimize both CPU and GPU performance.

\subsection{Optimization techniques}
\subsubsection{Precision.}
The investigated models support variable precision in computations. For our experiments, we utilized both double (64-bit), float (32-bit) and half (16-bit) precision. When reducing the precision of computations from double to float or half, several significant changes occur in the system's performance and resource utilization. The most immediate impact is a reduction in memory usage, as each value now occupies half the space. This reduction allows for larger batch sizes or higher resolution images to be processed simultaneously, enhancing the throughput of the model.

\subsubsection{Pin\_memory.}
During the execution of the algorithm, there is active communication and data exchange between the CPU and GPU. Among the investigated parameters \texttt{pin\_memory} showed positive effects on the algorithm's performance. This is a parameter of \texttt{DataLoader} \cite{doi:10.2352/EI.2024.36.12.HPCI-196} that forces the system to use only page-locked memory and prevents intermediate data from being swapped to disk. By locking the memory pages in RAM, it allows for faster and more efficient data transfers between the CPU and GPU.

\subsection{Selected profiler metrics}
As a result of the conducted tests, a comprehensive review of the metrics offered by the profiling tools was undertaken, leading to the selection of several key metrics. Below are some selected metrics that have the greatest impact on the performance of the tested applications.

%\begin{center}
\begin{table}[ht]
\caption{Selected profiler metrics used for profiling}
  \label{tab:metrics}\centering
\begin{tabular}{ | l | l | }
\hline
\thead{Metric} & \thead{Description} \\
\hline
\texttt{CUDA memcpy Host-to-Device} & \makecell[l]{time taken for data transfer from the CPU memory \\ to the GPU memory} \\
\hline
\texttt{cudaLaunchKernel} & \makecell[l]{time taken to launch a GPU kernel from the host}\\
\hline
\texttt{cudaStreamSynchronize} & \makecell[l]{function from the CUDA API designed to \\ synchronize task execution on the GPU} \\
\hline
\texttt{ncclDevKernel\_AllGather} & \makecell[l]{time taken by the NCCL to perform the All-Gather \\ operation using the Ring algorithm with the \\ Low-Latency protocol} \\
\hline
\texttt{cudaEventSynchronize} & \makecell[l]{function is used to synchronize CPU thread with \\ the GPU by forcing it to wait until all operations \\ linked to a particular event are fully completed} \\
\hline
\texttt{cudaEventDestroy} & \makecell[l]{function is employed to properly release the memory \\ and other resources associated with an event, once \\ it is no longer needed} \\ \hline
\texttt{cudaMemcpyAsync} & \makecell[l]{asynchronous memory copy function that enables \\ non-blocking data transfers between different \\ memory spaces, such as from CPU to GPU, \\ or between device memories} \\
\hline
\end{tabular}
\end{table}
%\end{center}

\section{Profiling of image recognition}
\label{sec:cpu}

\subsection{Benchmarking model}

\subsubsection{Mobilenet v2.}

\texttt{MobileNet v2} \cite{Simonyan2014VeryDC} is a convolutional neural network with a depth of 53 layers, specifically designed for efficient performance in resource-constrained environments, such as mobile devices. The architecture is characterized by an inverted residual structure, wherein the input and output of each residual block are narrow bottleneck layers, while channel expansion occurs within the intermediate layers. This design contrasts with traditional residual models, where expanded representations are utilized at the input and output. \texttt{MobileNet v2} also employs depthwise separable convolutions, significantly reducing the computational load and the number of parameters, while maintaining high accuracy.

In addition, the removal of non-linearities in the narrow bottleneck layers prevents information loss, enhancing the model's capacity for generalization. These architectural choices make \texttt{MobileNet v2} highly suitable for deployment in environments with limited computational resources. The model also supports adjustable width multipliers, allowing for scalability depending on the available computational power.

The implementation of \texttt{MobileNet v2} across several GPUs results in significant enhancements in both training and inference speeds, especially when processing extensive datasets. Its lightweight design and efficient use of parameters allow the model to scale proficiently with parallel computing, thereby maximizing the performance of multi-GPU configurations.

\subsection{Workload distribution strategies}

This chapter presents a thorough examination of strategies aimed at improving the performance of neural model training in the context of increased hardware capabilities. A range of data distribution methods was explored to emphasize their differing implementations and effectiveness. A method for achieving parallel execution is Data Parallelism, utilizing PyTorch's DataParallel mechanism in three different versions. This approach involves single-process multithreaded parallelism, which permits the same model to run concurrently on several GPUs. Its implementation is straightforward, as it does not necessitate additional code for configuring process groups, making it accessible for existing non-parallel projects.

\paragraph{DDP with no workload distribution (DDP).}
In this case all data sent to GPU is being processed. It is not being divided it into smaller chunks nor are there any other restrictions on processing applied. Since we are sending entire dataset to 4 accelerators, GPUs perform effectively 4 times the work.

\paragraph{DDP with Round-robin workload distribution (DDP-RR).}
This approach introduces a check inside parallelized process function that tests whether a part of dataset is assigned to a particular GPU. All data is still being sent to all GPUs, but in this case only a chunk of it is being processed by each of accelerators.

\paragraph{DDP with Distributed Sampler (DDP-DS).}
Distributed Data Sampler mechanism manages dividing the dataset into samples and parallelized workload distribution. This requires moving Data Loader logic into parallelized process function. However, thanks to it, only needed chunks of data are being sent to GPUs.

%The difference between the DDP-DS and DDP-RR algorithms is most evident in the Host-to-Device operations. DDP-DS minimizes the number of data transfers from host to device, as each process is assigned a unique subset of data, which reduces redundant data movement. According to profiler metrics, cudaMemcpyHostToDevice operations take less time and occur less frequently in DDP-DS, resulting in a shorter cumulative duration of these operations. In contrast, DDP-RR requires more frequent data transfers from host to device, as processes cyclically handle the same data, leading to increased time spent on Host-to-Device operations.

\hfill \break
Given the research focus of this study, the subsequent work will concentrate on the three most advanced distribution models: DDP, DDP-RR, and DDP-DS.

\subsection{Examined dataset and results quality}
\subsubsection{\texttt{MNIST} dataset.}
To evaluate different approaches to GPU load distribution, it is crucial to ensure that our interventions do not affect the algorithm's outcomes. This requires stable input data and a quantifiable metric for neural network training quality. \texttt{MNIST} is a dataset containing monochrome handwritten digit images, each sized 28x28 pixels, with 60,000 training images and 10,000 test images. After performing several tests, we learned that the 28x28 pixel size was too small to significantly load the GPU (memory management and communication operations predominated over computation). Therefore, the dimensions of the analysis output images were modified from 100x100 to 500x500 pixels in order to evaluate the impact of this alteration on the GPU load.

\subsubsection{Model training quality.}
In the conducted research, we analysed both the effects of different parameters on the model's runtime and their impact on computational accuracy. In conclusion, the primary factor for attaining consistent and satisfactory outcomes was identified as the number of training epochs. Our objective was to achieve an accuracy rate of 98-99\% on the test dataset, which generally necessitated 20 training epochs. It was observed that enlarging the image size resulted in an extended model runtime, yet it produced more precise results.

\subsection{Evaluation of findings for image recognition}

\subsubsection{Data distribution results}
\begin{figure}[ht]
    % \centering
    \includegraphics[width=1\linewidth]{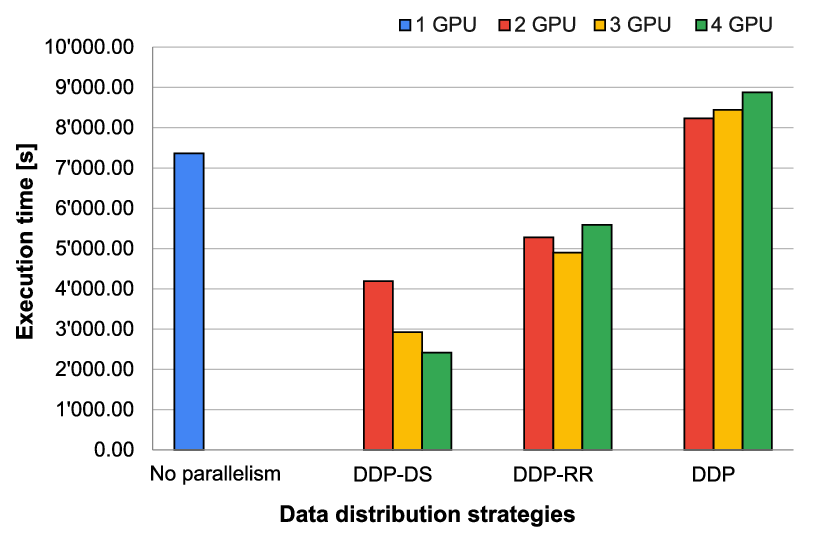}
    \caption{Distributed learning algorithm performance scaling, \texttt{FP32}, $pin\_memory = FALSE$}
    \label{fig:ddp_algos_diagram}
\end{figure}

The Figure \ref{fig:ddp_algos_diagram} presents the performance testing results of distributed learning algorithms using 1, 2, 3, and 4 GPUs with various data distribution strategies: DDP-DS, DDP-RR, and DDP. The experiments were conducted using \texttt{FP32} numbers with pinned memory disabled and maximum possible picture size 700x700. The results indicate that the DDP-DS algorithm effectively scales with the addition of new computational resources, while DDP-RR and DDP show limited or no performance gains. Therefore, further tests will focus on comparing the performance of a single GPU with the DDP-DS approach.

\begin{figure}[ht]
    % \begin{sidewaysfigure}
        % \centering
        \includegraphics[width=0.49\linewidth]{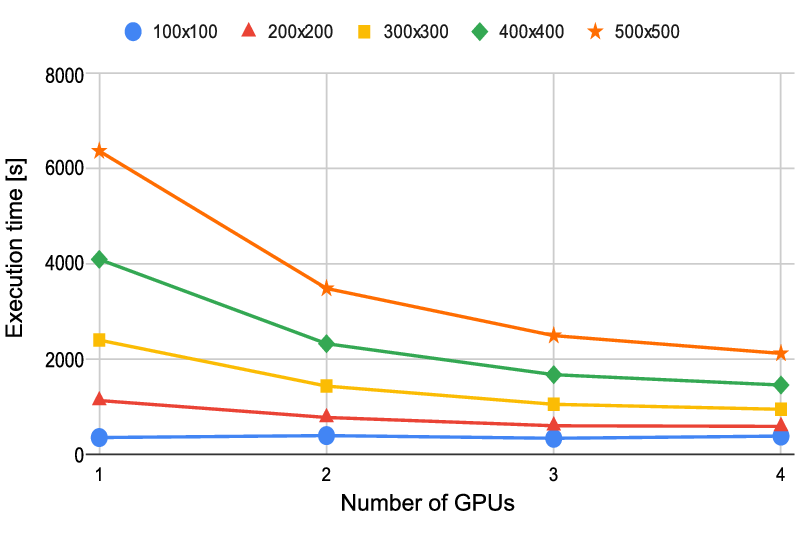}
        \includegraphics[width=0.49\linewidth]{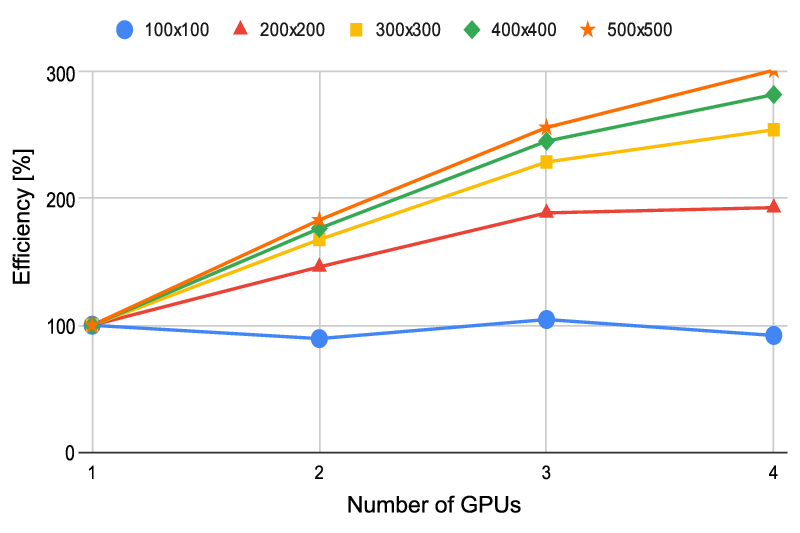}
    % \end{sidewaysfigure}
    \caption{Execution time (left) and efficiency (right) graphs for image sizes from 100x100 to 500x500, \texttt{FP64}, $pin\_memory = false$}
    \label{fig:diff_sizes_graph}
\end{figure}

The Figure \ref{fig:diff_sizes_graph} presents the performance results of the DDP-DS algorithm for image sizes ranging from 100x100 pixels to 500x500 pixels. From these graphs, it is evident that the benefits of adding new GPUs become noticeable only at a certain image size that can create sufficient load on the computational power of the GPUs. The graph shows that effective utilization of 4 GPUs is observed with image sizes starting from 300x300 pixels.

\subsubsection{Optimization results}

\paragraph{Precision.}
Switching from double (\texttt{FP64}) to float (\texttt{FP32}) or half (\texttt{FP16}) precision yields a significant improvement in all presented cases. In table \ref{tab:precision_table} it can be observed a consistent increase in computational speed by 54-68\% for float and 110-152\% for half compared to using the same parameter set with double precision. Additionally, as shown in the 'Success test dataset' column, the quality of results does not deteriorate when transitioning from double to float or half precision. This can be attributed to the fact that training is performed over 20 epochs, during which the network achieves satisfactory training quality.

\begin{table}[ht]
    \caption{Comparison of algorithm performance for \texttt{FP16}, \texttt{FP32} and \texttt{FP64}, $pin\_memory=FALSE$}
    \label{tab:precision_table}
    \centering
    \adjustbox{max width=\linewidth}{
    \begin{tabular}{c c c c c c}
        \toprule
        \textbf{GPUs} & \textbf{Precision} & \makecell{\textbf{Total} \\ \textbf{time [s]}} & \makecell{\textbf{Inference} \\ \textbf{accuracy} \\ \textbf{[\%]}} & \makecell{\textbf{Efficiency} \\ \textbf{compare} \\ \textbf{to 1 GPU}} & \makecell{\textbf{Efficiency} \\ \textbf{compare} \\ \textbf{to \texttt{FP64}}} \\
        \midrule
        % 1 & FP64 & 6\,389 & 99.30 & 100.00\% & -       \\
        % 1 & FP32  & 3\,792 & 99.30 & 100.00\% & 168.51\% \\
        % 2 & FP64 & 3\,450 & 99.36 & 185.21\% & -       \\
        % 2 & FP32  & 2\,205 & 99.24 & 171.93\% & 156.42\% \\
        % 3 & FP64 & 2\,472 & 99.12 & 258.39\% & -       \\
        % 3 & FP32  & 1\,579 & 99.30 & 240.01\% & 156.52\% \\
        % 4 & FP64 & 2\,110 & 99.12 & 302.80\% & -       \\
        % 4 & FP32  & 1\,358 & 99.20 & 279.08\% & 155.31\% \\
        % \\
        1 & FP64 & 6\,266 & 99.43 & 100\% & -       \\
        1 & FP32  & 3\,721 & 99.32 & 100\% & 168.38\% \\
        1 & FP16  & 2\,477 & 99.57 & 100\% & 252.93\% \\
        2 & FP64 & 3\,339 & 99.20 & 187.66\% & -       \\
        2 & FP32  & 2\,094 & 99.24 & 177.69\% & 159.45\% \\
        2 & FP16  & 1\,445 & 97.94 & 171.41\% & 231.07\% \\
        3 & FP64 & 2\,318 & 98.70 & 270.31\% & -       \\
        3 & FP32  & 1\,483 & 99.03 & 250.91\% & 156.30\% \\
        3 & FP16  & 1\,049 & 99.21 & 236.12\% & 220.97\% \\
        4 & FP64 & 1\,937 & 99.32 & 323.48\% & -       \\
        4 & FP32  & 1\,256 & 98.88 & 296.25\% & 154.21\% \\
        4 & FP16  & 919 & 99.20 & 269.53\% & 210.77\% \\
        \bottomrule
    \end{tabular}}
\end{table}

Upon examining the profiler and comparing changes in various metrics, we can observe that the largest reduction in time occurs at the [CUDA memcpy Host-to-Device] stage and \texttt{cudaMemcopyAsync}. The Host-to-Device copy time decreases proportionally with a twofold reduction in precision for float, and a fourfold reduction for half precision, from 226s for \texttt{FP64} to 104s for \texttt{FP32} and 50s for \texttt{FP16}. \texttt{cudaMemcopyAsync} also decreases proportionally: 1\,334s for \texttt{FP64}, 724s for \texttt{FP32} and 478s for \texttt{FP16}. This is the most significant change. Among other metrics, there are also changes in $ncclDevKernel\_Broadcast$, where the time decreases proportionally from 40s for \texttt{FP64} to 27s for \texttt{FP32} and 15s for \texttt{FP16}. Among the remaining metrics, it is also worth noting the \texttt{cudaEventDestroy} metric, whose execution time decreases threefold or even more, rather than twofold, with each reduction in precision, from 205s for \texttt{FP64} to 84s for \texttt{FP32} and 16s for \texttt{FP16}.

\paragraph{$pin\_memory$.}
Table \ref{tab:pinMemory_table} presents a performance improvement from using \break $pin\_memory$, ranging between 16\% and 30\%, with the efficiency decreasing as the number of GPUs increases.

\begin{table}[ht]
    \caption{Comparison of algorithm performance for $pin\_memory$ flag, precision \texttt{FP32}}
    \label{tab:pinMemory_table}
    \centering
    \adjustbox{max width=\linewidth}{
    \begin{tabular}{c c c c c c}
        \toprule
        \textbf{GPUs} & \makecell{\textbf{Pin} \\ \textbf{memory}} & \makecell{\textbf{Total} \\ \textbf{time [s]}} & \makecell{\textbf{Inference} \\ \textbf{accuracy} \\ \textbf{[\%]}} & \makecell{\textbf{Efficiency} \\ \textbf{compare} \\ \textbf{to 1 GPU}}  & \makecell{\textbf{Efficiency} \\ \textbf{compare} \\ \textbf{to $FALSE$}} \\
        \midrule
        1 & FALSE & 3\,792 & 99.30 & 100.00\% & -       \\
        1 & TRUE  & 2\,921 & 99.41 & 100.00\% & 129.78\% \\
        2 & FALSE & 2\,205 & 99.24 & 171.93\% & -       \\
        2 & TRUE  & 1\,733 & 99.48 & 168.56\% & 127.23\% \\
        3 & FALSE & 1\,579 & 99.30 & 240.01\% & -       \\
        3 & TRUE  & 1\,283 & 98.70 & 227.57\% & 123.05\% \\
        4 & FALSE & 1\,358 & 99.20 & 279.08\% & -       \\
        4 & TRUE  & 1\,165 & 99.36 & 250.79\% & 116.62\% \\
        \bottomrule
    \end{tabular}}
\end{table}

% Wnioski
The analysis shows that switching to lower precision (\texttt{FP32} and \texttt{FP16}) from double precision (\texttt{FP64}) provides significant performance improvements, with up to 210\% speed increase for \texttt{FP16} without sacrificing inference accuracy. Using pinned memory ($pin\_memory$) further enhances performance by 16-30\%, though the efficiency gain decreases as the number of GPUs increases. Profiling reveals that the most substantial time reduction occurs during CUDA Host-to-Device transfers and \texttt{cudaMemcopyAsync}, with \texttt{FP16} achieving the lowest times. However, increasing the number of GPUs and enabling $pin\_memory$ also leads to increased execution times for NCCL kernels, diminishing some of the benefits obtained from faster \texttt{Host-to-Device} communication. Overall, reduced precision and optimized memory handling are crucial in enhancing computational efficiency in multi-GPU environment.

\subsubsection{Metric analysis}
%https://docs.google.com/spreadsheets/d/1asHK7kXXzWujGfDK1TUrmDF9pe3zW6H4d2Z4jOMNwm8/edit?usp=sharing - link do danych
\hfill \break
While analyzing the data provided by the profiler, we found that these optimization techniques affect the execution time of different kernels and memory operations. We selected four of the most interesting metrics: \texttt{CUDA memcpy Host-to-Device}, \texttt{cudaLaunchKernel}, \texttt{cudaStreamSynchronize}, and \hfill \break \texttt{ncclDevKernel\_AllGather\_RING\_LL}.

\begin{figure}[ht]
    % \centering
    \includegraphics[width=0.9\linewidth]{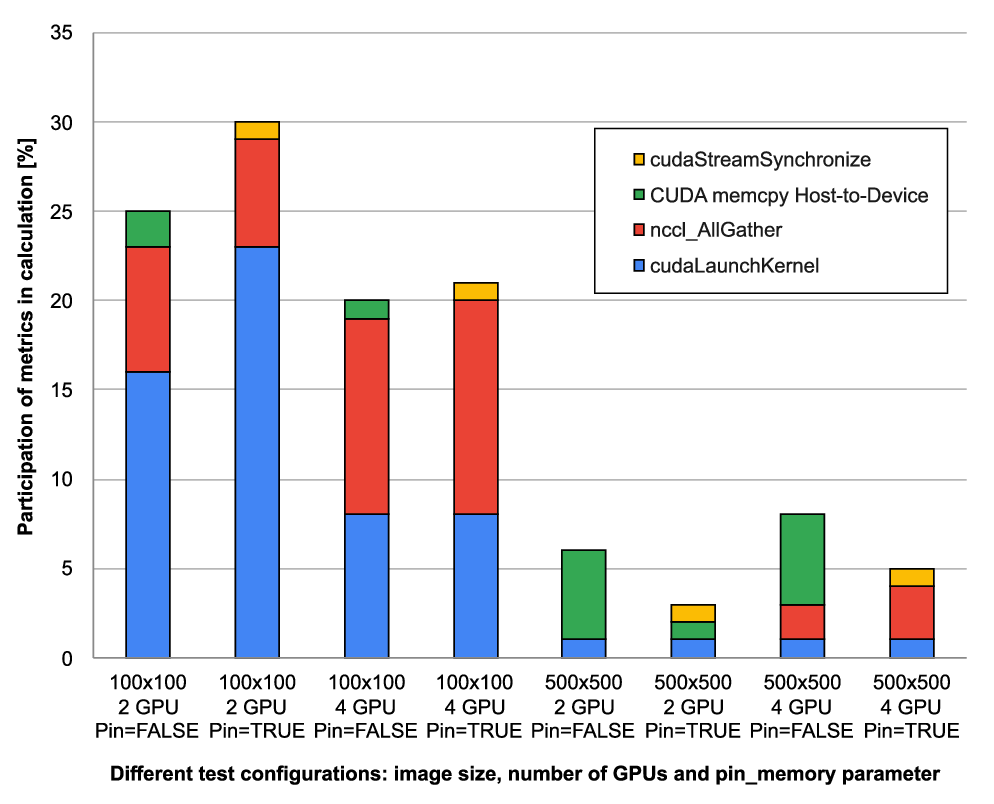}
    \caption{Diagram with 4 metrics times from NSight System in percent}
    \label{fig:metric_percent_diagram}
\end{figure}

In Figure \ref{fig:metric_percent_diagram} we present 8 tests for \texttt{FP32} with 3 input parameter variations: image size (100x100 or 500x500), number of GPUs (2 or 4), and $pin\_memory$ (true or false). We observe that \texttt{cudaLaunchKernel} and \texttt{ncclDevKernel\_AllGather\_RING\_LL} take relatively more time when the image size is small. This indicates that the algorithm spends more time on computation management and distribution than on the computations themselves. Additionally, it can be seen that the time for \texttt{CUDA memcpy Host-to-Device} significantly decreases when $pin\_memory$ is enabled, but the time spent on \texttt{cudaStreamSynchronize} increases.

\section{Profiling of Large Language Models}

In this part of the study, we evaluate the performance of LLAMA3-8B LoRA \cite{10695440} fine-tuning process using differing datasets in a multi-GPU environment. Additionally data transfer/communication operations are also explored in this context. Finally, a comparison of these approaches and conclusions are provided.

\subsection{Model specification}

LLama3 is a large language model developed by Meta trained to perform tasks such as text generation, translation, question answering, and sentiment analysis \cite{minaee2024largelanguagemodelssurvey}. It utilizes transformer architecture with Grouped-Query Attention  to improve inference efficiency. Training was performed on the internet-scale corpus of diverse texts of over 15T tokens. It uses a tokenizer with a vocabulary of 128K tokens, and its context length (the number of tokens considered by the model to predict the next word) is set at 8,192 tokens. The LLAMA3-8B variant of Llama consists of 8 billion parameters, with March, 2023 knowledge cutoff date. Smaller size makes it less memory demanding and easier to deploy in various applications. The model can be fine-tuned on specific tasks by further training it on task-specific data, allowing it to perform better on those tasks.

\subsection{LLM fine-tuning process}

Fine-tuning is a process separate from model training, that can help the model improve its performance on particular tasks where additional knowledge is required beyond what can be learned from general text data alone.

\subsubsection{Tuning techniques.}

There are several approaches to tuning LLMs. Because of large number of parameters present in these models, efficient approaches are necessary. One example is LoRA (Low-Rank Adaptation) - a technique involving attaching a small number of low-rank adapter layers to a pre-trained LLM. These are trained on specific downstream tasks and improve performance without modifying the model's architecture \cite{hu2021loralowrankadaptationlarge}. LoRA approach reduces the computational overhead and memory requirements of fine-tuning large language models.

In contrast, Direct Preference Optimization (DPO) directly optimizes the LLM parameters to maximize the likelihood of generating text that is closest to provided preference data. It simplifies tuning by not requiring a additional, external steps, like separate reward model or Reinforcement Learning from Human Feedback \cite{rafailov2023direct}.

Further optimization to LoRA is quantized low rank adaptation (QLoRA), which additionally utilizes quantization mechanism (compressing model parameters to a 4-bit format) to further reduce VRAM requirements at the cost of increased compute time and slight accuracy loss \cite{dettmers2023qloraefficientfinetuningquantized}.

The choice between LoRA and DPO for fine-tuning LLMs often depends on specific goals and constraints of the application. For scenarios where computational efficiency is crucial but flexibility isn't paramount, LoRA could be a preferred method due to its efficient parameter usage. As computing performance is the main focus of this work, it has been chosen for LLM tests.

\subsubsection{Methodology for LLMs fine-tuning.}

The actual tuning was conducted utilizing TorchTune using the recipes for single-device and distributed Llama3-8B LoRA. These recipes were subsequently modified using custom configuration files in order to perform necessary test variants. These configuration files contain parameters that control and optimize the tuning process. For the following parameters, the default values were used - batch size: 2, learning rate: 3e-4, warm-up steps: 100, epochs: 1.

\subsection{Investigated datasets} \label{subsec:datasets}
In order to properly understand the performance impact of tuning, it is first necessary to understand the differences between datasets used. Here we present details of a selection of datasets from \textit{Torchtune} library. Each one uses a particular template, which are used to format prompts to optimize model performance on specific tasks, e.g. answering questions, summarizing or correcting errors. Each template includes the template prompt with placeholders for the data inputs.

\subsubsection{alpaca-cleaned:}
This is a cleaned version of the original Alpaca Dataset released by Stanford, which was generated by a language model \textit{text-davinci-003}.
This instruction data is designed for instruction-tuning for pertrained language models. The tasks consist of answering questions in a concise manner, using built-in knowledge and reasoning, including classification, instruction following, and writing. The template used for tuning is InstructTemplate.

\subsubsection{C4\_200M Synthetic Dataset for Grammatical Error Correction (grammar)}
It is the largest of tested datasets and can be used in grammatical error correction (GEC) tasks. As evidenced, the tasks assigned include finding words or phrases that contain language errors and replace them with correct form.

\subsubsection{SAMsum}
The \texttt{SAMSum} dataset is the smallest of tested datasets and was created to emulate conversation styles and topics commonly encountered in messaging apps. Each conversation includes a unique identifier and detailed metadata such as dialogue text, summary, speaker names, and more. Conversations vary in formality, containing diverse language including slang, emoticons, and typos.

\subsubsection{SlimOrca Dedup.}
\texttt{SlimOrca Dedup (slimorca)} is a deduplicated, unfiltered subset of the subset of  OpenOrca data, which comprises entries from the FLAN Collection, augmented by querying GPT-4 or GPT-3.5 API with specific questions to elicit detailed reasoning responses. Supported tasks include language modeling, text generation, and augmentation. The template used for tuning is InstructTemplate, same as Alpaca dataset.
%m{6cm}|
\begin{table}[ht]
\centering
\caption{Dataset characteristics}
\label{tab:dataset_summary}
{\renewcommand{\arraystretch}{1.4}
\begin{tabular}{|l l l m{3.5cm} m{2.5cm}|}
\hline
Dataset & Size (MB) & Length (pairs) & Tasks & Template\\ \hline
Alpaca cleaned & 24.1 & 51,760 & Classification, summarization, and writing & Instruct \\  \hline
Grammar & 3710.5 & 185,000,000 & Correcting grammatical errors & Grammar Error Correction  \\  \hline
Samsum & 10.71 & 16,369 & Conversation summarization & Summarize   \\  \hline
SlimOrca & 307 & 363,491 & Language modeling, text generation, and augmentation & Instruct  \\ \hline
\end{tabular}}
\end{table}
\vspace{10pt}

\subsection{Evaluation}

All tests in this section were performed using the same value for the following parameters: batch size:\textit{2}, learning rate: $3\mathrm{e}{-4}$, warm-up steps: \texttt{100}, epochs: \texttt{1}. Detailed information is provided as description to each relevant test.

\subsubsection{Optimization results.}
\begin{figure}[ht]
%źródło: arkusz - LLM/Energy, memory, iteration time results; skoroszyt - wykres
    \centering
    \includegraphics[width=1\linewidth]{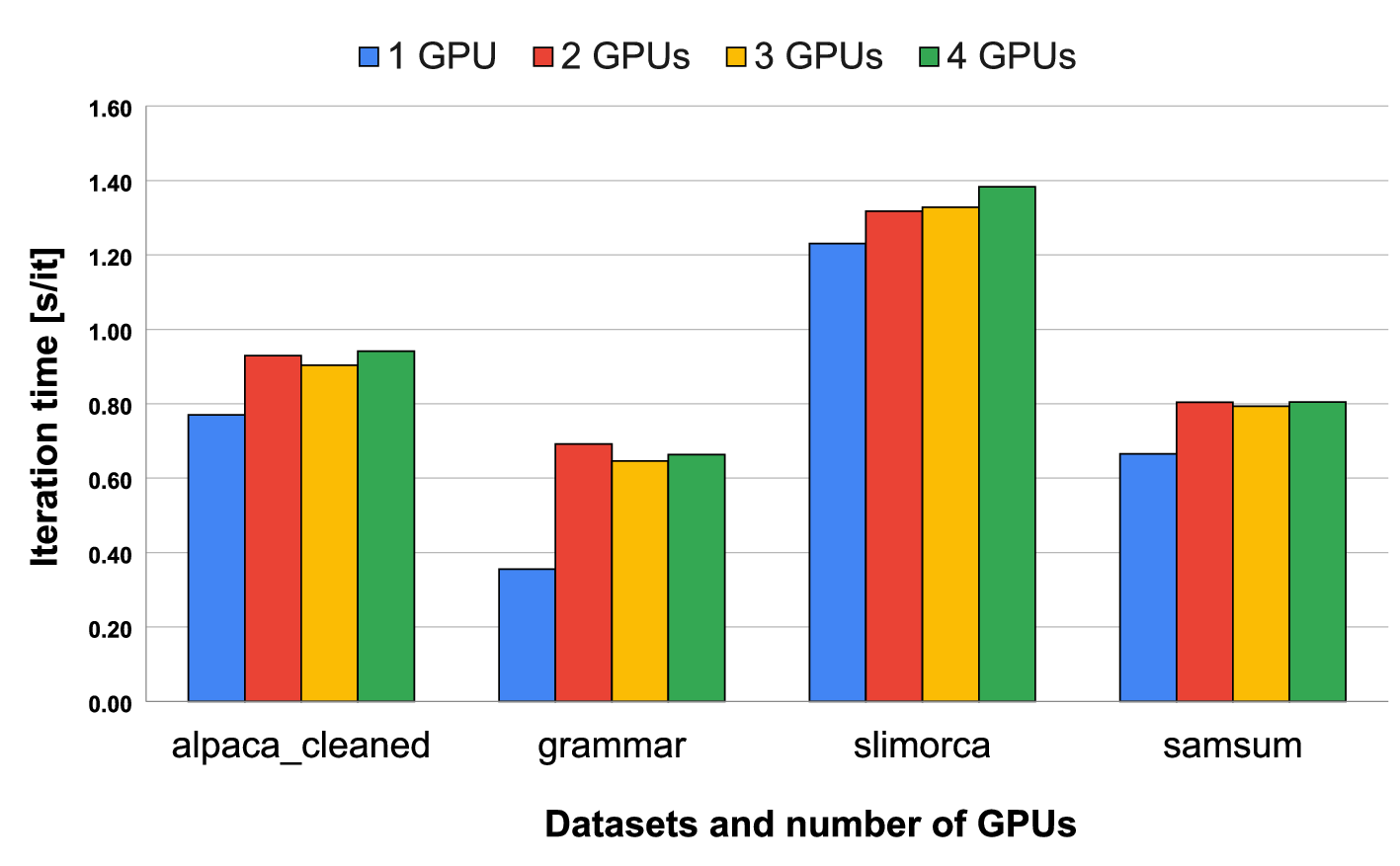}
    \caption{Dataset tuning performance scaling}
    \label{fig:llmiterationtime}
\end{figure}

Performance impact of dataset being tuned was measured by determining mean time it takes to complete an iteration - a group of dataset rows processed concurrently, with its size dependent on number of GPUs and batch size.
Figure \ref{fig:llmiterationtime} depicts average iteration times for LLM tuning for specified datasets: \texttt{alpaca-cleaned}, \texttt{grammar}, \texttt{samsum} and \texttt{slimorca}, compared when scaling using 1, 2, 3, and 4 GPUs. The experiments were conducted using \texttt{FP32} precision with pinned memory option disabled.

The results indicate that there is a slight increase in average iteration time when adding more GPUs for all datasets. This can be explained by additional communication and synchronization between GPUs needed for finishing each iteration. \texttt{Grammar} and \texttt{samsum} appear to have shorter iteration time, whereas \texttt{slimorca} has the longest. In addition adding more than 2 GPUs appears to have little effect on iteration time. Of course, utilizing more GPUs reduces the number of iterations necessary to process the whole dataset, which in turn cuts total time to fine-tune. All runs testing \texttt{$pin\_memory$} flag were executed on \texttt{alpaca-cleaned} dataset with \texttt{FP32} precision and \textit{batch size} set to 2.

\begin{table}[ht]
%źródło: arkusz - LLM/Optimization results
    \centering
    %\adjustbox{max width=\linewidth}{
    \begin{tabular}{ccccccccc}
        \toprule
        \textbf{GPUs} & \textbf{$pin\_memory$} & \textbf{Iteration time (s/it)} & \textbf{Total time} & \textbf{Relative improvement} \\
        \midrule
        1 & FALSE & 0.77 & 5:31:52 & -       \\
        1 & TRUE  & 0.77 & 5:31:44 & 0,04\% \\
        2 & FALSE & 0.93 & 3:20:30 & -       \\
        2 & TRUE  & 0.92 & 3:17:21 & 1,57\% \\
        3 & FALSE & 0.90 & 2:09:55 & -       \\
        3 & TRUE  & 0.90 & 2:09:57 & -0,03\% \\
        4 & FALSE & 0.94 & 1:41:32 & -       \\
        4 & TRUE  & 0.91 & 1:37:58 & 3,50\% \\
    \end{tabular}%}
    \vspace{5pt}
    \caption{Comparison of $pin\_memory$ option performance impact.}
    \label{tab:pinmemorytime}
\end{table}

The table \ref{tab:pinmemorytime} presents a comparison between tuning time run on 1, 2, 3 or 4 GPUs with and without $pin\_memory$ flag enabled. There is very little variation in execution time between cases with and without this flag present. The performance gains of this option appear to be negligible in case of LLM tuning.

\subsubsection{Metric analysis.}
Next series of tests utilize profiler data, in order to measure operations that collectively take up the most computing time - \texttt{cudaLaunchKernel} and \texttt{cudaStreamSynchronize} and memory operations \texttt{Host-to-Device}.

The tests were performed on different datasets, using a single GPU and \texttt{FP32} precision. Because tuning process can take up considerable time, which is also variable between datasets, in order to make a just comparison, profiling results were collected after first 12 minutes (720s) of processing in all cases. Additionally, \texttt{nccldev\_AllGather} metric is not included, as it is only executed during distributed tuning (using more than 1 GPU).

\iffalse
\begin{table}[ht]
%źródło: arkusz - LLM/nsys; skoroszyty - proxima alpaca, proxima grammar, proxima slimorca, proxima samsum
    \centering
    \adjustbox{max width=\linewidth}{
    \begin{tabular}{lcccc}
        \toprule
        \textbf{Dataset} & \textbf{API call} & \textbf{Time(\%)} & \textbf{Total Time (s)} & \textbf{Iteration time (s/it)}\\
        \midrule
        alpaca-cleaned & cudaLaunchKernel & 19.8\% & 227.973 & 0.77 \\
        grammar & cudaLaunchKernel  & 13.6\% & 134.447 & 0.36 \\
        slimorca & cudaLaunchKernel & 20.9\%  & 246.633 & 1.23 \\
        samsum & cudaLaunchKernel  & 18.9\%  & 216.052 & 0.67 \\
        alpaca-cleaned & cudaStreamSynchronize & 23.30\% & 268.144 & 0.77 \\
        grammar & cudaStreamSynchronize  & 25.30\% & 250.179 & 0.36 \\
        slimorca & cudaStreamSynchronize & 21.0\% & 248.012 & 1.23 \\
        samsum & cudaStreamSynchronize  & 23.8\% & 271.527 & 0.67 \\
        alpaca-cleaned & Host-to-Device & 0.10\% & 1.054 & 0.77 \\
        grammar & Host-to-Device  & 0.10\% & 1.054 & 0.36 \\
        slimorca & Host-to-Device & 0.10\% & 1.049 & 1.23 \\
        samsum & Host-to-Device  & 0.10\% & 1.057 & 0.67 \\
    \end{tabular}}
    \vspace{5pt}
    \caption{Comparing \texttt{cudaLaunchKernel} and \texttt{cudaStreamSynchronize} time for different datasets.}
    \label{tab:datasetapi}
\end{table}
\fi

\begin{figure}
%źródło: arkusz - LLM/nsys; skoroszyt - wykres dataset
    %\centering
    \includegraphics[width=0.5\linewidth]{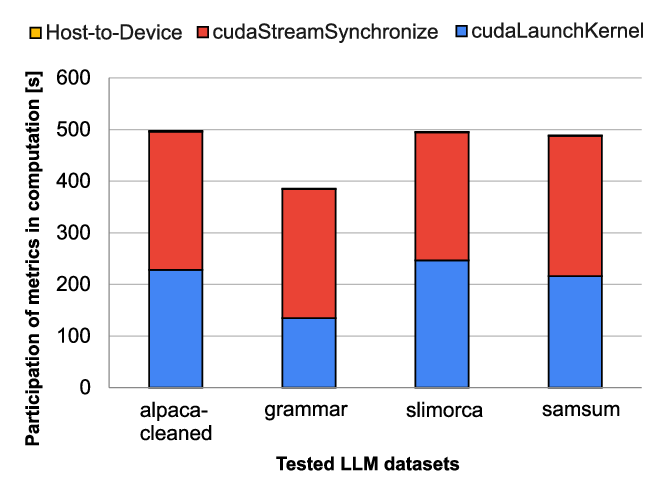}
    \includegraphics[width=0.5\linewidth]{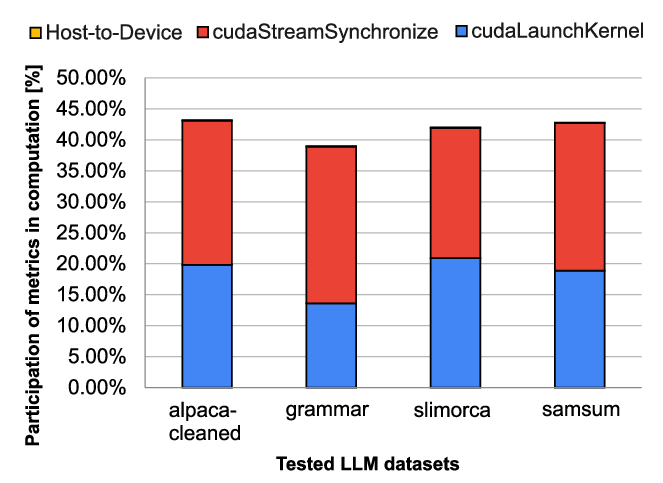}
    \caption{Cuda API calls time for tested datasets}
    \label{fig:llmdatasetapi}
\end{figure}

As can be noted in figure \ref{fig:llmdatasetapi} the general trend appears to be that significant amount of processing time is spent on launching new kernels and synchronizing numerous threads. This can be interpreted by the multitude of iterations, that need to be sequentially executed in order to complete the tuning process. Additionally, total time spent on executing \texttt{cudaLaunchKernel} operations seems to correlate with average iteration time of the dataset. Conversely, with \texttt{cudaStreamSynchronize} calls, there is much less variation in total time and no such correlation occurs. Therefore, we conclude that communication operations remain relatively constant on a single GPU, regardless of dataset size or its tasks, and time spent on launching new kernels is related to the speed of iteration processing. Memory operations do not seem to have a significant impact on processing time in any case.

%$pin\_memory$ ma wpływ na operacje pamięciowe, ale nie ma na całkowity czas cudastreamsynchronize, bo to dzieje się na gpu, a $pin\_memory$ dotyczy tylko pamięci

For the API call tests for $pin\_memory$ flag, analogously as in case of dataset measurements, profiling results were collected after first 12 minutes of processing.

\iffalse
\begin{table}[ht]
%źródło: arkusz - LLM/nsys; skoroszyty pin[1-2]report i alpaca[1-2]
    \centering
    \begin{tabular}{clcc}
        \toprule
        \textbf{GPUs} & API call & \textbf{$pin\_memory$} & \textbf{Total Time(s)} \\
        \midrule
        1 & Host-to-Device & TRUE & 1.400     \\
        1 & Host-to-Device & FALSE & 1.054   \\
        \hline
        2 & Host-to-Device & TRUE & 1.090     \\
        2 & Host-to-Device & FALSE & 1.056   \\
        \hline
        1 & cudaLaunchKernel & TRUE & 246.106     \\
        1 & cudaLaunchKernel & FALSE & 247.708   \\
        \hline
        2 & cudaLaunchKernel & TRUE & 41.775    \\
        2 & cudaLaunchKernel & FALSE & 39.597   \\
        \hline
        1 & cudaStreamSynchronize & TRUE & 288.591 \\
        1 & cudaStreamSynchronize & FALSE & 291.463 \\
        \hline
        2 & cudaStreamSynchronize & TRUE & 121.193 \\
        2 & cudaStreamSynchronize & FALSE & 120.860 \\
        \hline
        2 & nccldev\_AllGather & TRUE & 653.629     \\
        2 & nccldev\_AllGather & FALSE & 648.541  \\
    \end{tabular}
    \vspace{5pt}
    \caption{Comparing CUDA \texttt{Host-to-Device} memory operations, \texttt{nccldev\_AllGather}, \texttt{cudaLaunchKernel} and \texttt{cudaStreamSynchronize} time with \texttt{pin\_memory} option impact.}
    \label{tab:pinmemoryapi}
\end{table}
\fi

\begin{figure}[ht]
%źródło: arkusz - LLM/nsys; skoroszyt - pin wykres
    %\centering
    \includegraphics[width=0.5\linewidth]{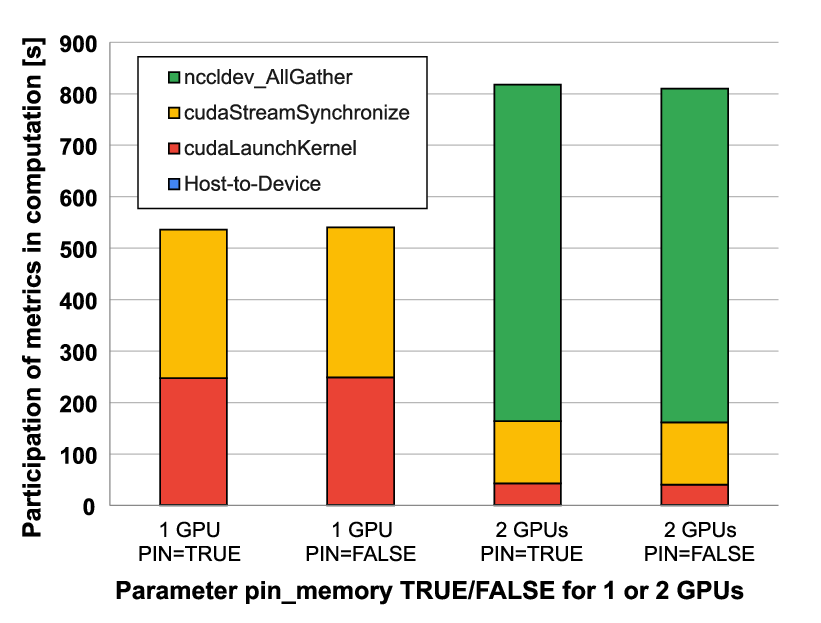}
    \includegraphics[width=0.5\linewidth]{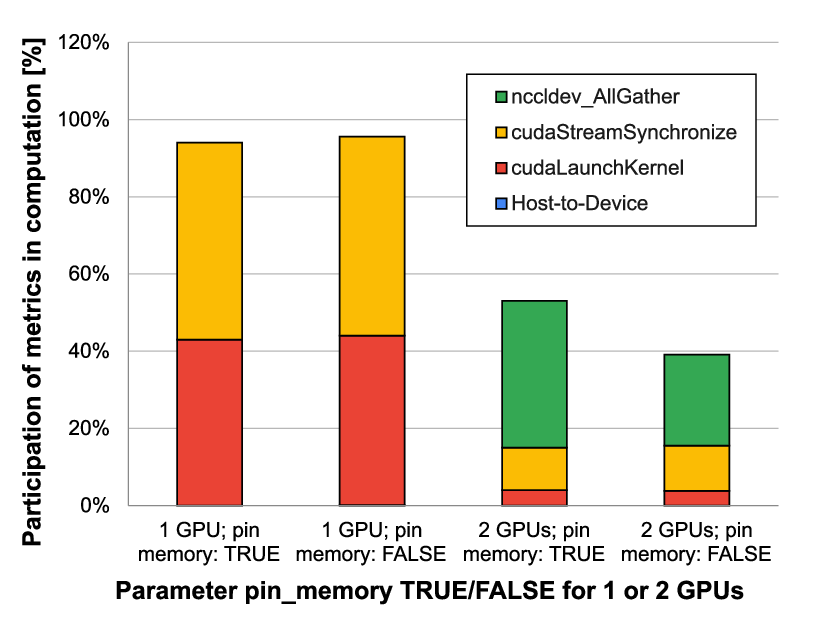}
    \caption{Cuda API calls time for \texttt{pin\_memory} in seconds (left) and percent (right) distributed on 1 or 2 GPUs}
    \label{fig:llmpinmemoryapi}
\end{figure}

Chart displayed in Figure \ref{fig:llmpinmemoryapi} presents profiling data from API calls with respect to \texttt{pin memory} option. In addition to \texttt{cudaLaunchKernel}, \texttt{Host-to-Device} memory transfers were measured, as well as \texttt{nccldev\_AllGather} in cases where tuning was distributed among multiple GPUs. Because profiling data was gathered from all accelerators, due to concurrency it is possible for total sum to exceed 720s.

With \texttt{pin\_memory} enabled there appears to be a slight increase in already insignificant Host-to-Device memory operations. Apart from that, there are no apparent performance gains from using this optimization technique.

In conclusion, with modern hardware it appears even the largest datasets do not seem to generate enough memory operations to influence tuning process much. The limited number of memory operations can be explained by small dataset size relative to model size and performing training with only one epoch, which means each row in dataset is only used once.

%- podczas strojenia brak komunikacji z CPU
%- mniejszy wpływ Pin memory w LLM niż dla rozpoznawania obrazów

%%%%%%%%%%%%%%%%%%%%%%%%%%%%%%%%%%%%%%%%%%%%%%%%%%

\section{Conclusions}

In this work authors elaborated efficiency of machine learning and LLM models utilization at multi-GPU cards system. Two optimization strategies, namely precision reduction and the use of $pin\_memory$, were proposed and evaluated, with a thorough analysis of the operational costs conducted using specific metrics.

In the context of the image recognition machine learning model (MobileNet v2), the analysis indicated that diminishing the precision of calculations (from FP64 to FP32 and then to FP16) can lead to a remarkable enhancement in computational speed, achieving increases of up to $210\%$ without sacrificing inference accuracy. Additionally, the implementation of $pin\_memory$ contributed to a further performance boost ranging from $16\%$ to $30\%$. However, that the performance gains diminish as the number of GPUs increases. The profiling process indicated that the most significant reduction in time occurs during the \texttt{CUDA Host-to-Device} and \texttt{cudaMemcpyAsync} transfers, with \texttt{FP16} demonstrating the shortest transfer times. Nevertheless, as the number of GPUs rises and $pin\_memory$ is activated, there is a tendency for longer execution times in NCCL kernels, which mitigates some of the advantages gained from accelerated \texttt{Host-to-Device} communication. In summary, the reduction of precision and the optimization of memory handling are essential for enhancing computational efficiency in a multi-GPU setting for the \texttt{MobileNet} model.

In the subsequent use case examined (LLM model tuning), a slight increase in the average iteration time was noted when incorporating additional GPUs (up to a maximum of two) across all datasets. Throughout the tuning process, the communication operations remain relatively stable on a single GPU, irrespective of the dataset size or the tasks involved, while the duration required to initiate new kernels is associated with the speed of iteration processing. A considerable amount of the processing time during LLM tuning is attributed to initiating new kernels and synchronizing multiple threads. Memory operations do not significantly contribute to the overall time spent on LLM tuning. The performance improvement attributed to \texttt{pin\_memory} appears to be minimal in the context of LLM tuning.

The scope of future work includes conducting tests with higher load (larger individual specimens in respective datasets), identifying new key metrics that affect performance and searching for further universal optimization techniques such as using the DALI (NVIDIA Data Loading Library) library.

% \subsection{Recommendations for utilizing data and hardware optimally}
% ---Image recognition
% In summary, the results indicate that reducing precision to float or half precision can significantly improve computational speed without compromising accuracy, and even reduce the time to minimum loss, making it a viable strategy for optimizing training performance. Moreover, while pinned memory helps improve data transfer efficiency, its impact on overall training convergence is minimal, emphasizing that its main benefit lies in improving data processing efficiency rather than in influencing learning dynamics.

% ---LLM
% ...

\section*{Acknowledgements}
Funded by the European Union. This work has received funding from the European High Performance Computing Joint Undertaking and Poland, Germany, Spain, Hungary, France and Greece under grant agreement number: 101093457. This publication expresses the opinions of the authors and not necessarily those of the EuroHPC JU and Associated Countries which are not responsible for any use of the information contained in this publication.

The results presented in this study were prepared using the infrastructure of the Poznan Supercomputing and Networking Center.

\bibliographystyle{plain}
\bibliography{main}

\begin{thebibliography}{10}

\bibitem{lin2024universalperformancemodelingmachine}
Zhongyi Lin, Ning Sun, Pallab Bhattacharya, Xizhou Feng, Louis Feng, and John~D. Owens.
\newblock Towards universal performance modeling for machine learning training on multi-gpu platforms, 2024.

\bibitem{10.1145/1058129.1058148}
K.~Fatahalian, J.~Sugerman, and P.~Hanrahan.
\newblock Understanding the efficiency of gpu algorithms for matrix-matrix multiplication.
\newblock In {\em Proceedings of the ACM SIGGRAPH/EUROGRAPHICS Conference on Graphics Hardware}, HWWS '04, page 133–137, New York, NY, USA, 2004. Association for Computing Machinery.

\bibitem{TorchtuneConfigs}
The~Linux Foundation.
\newblock {Torchtune config documentation}, 2024.

\bibitem{doi:10.2352/EI.2024.36.12.HPCI-196}
Edgar~Josafat Martinez-Noriega, Chen Peng, and Rio Yokota.
\newblock High-performance data loader for large-scale data processing.
\newblock {\em Electronic Imaging}, 36(12):196--1--196--1, 2024.

\bibitem{Simonyan2014VeryDC}
Karen Simonyan and Andrew Zisserman.
\newblock Very deep convolutional networks for large-scale image recognition.
\newblock {\em CoRR}, abs/1409.1556, 2014.

\bibitem{10695440}
Hongyi Shui, Yuanjing Zhu, Fan Zhuo, Yibo Sun, and Daoyuan Li.
\newblock An emotion text classification model based on llama3-8b using lora technique.
\newblock In {\em 2024 7th International Conference on Computer Information Science and Application Technology (CISAT)}, pages 380--383, 2024.

\bibitem{minaee2024largelanguagemodelssurvey}
Shervin Minaee, Tomas Mikolov, Narjes Nikzad, Meysam Chenaghlu, Richard Socher, Xavier Amatriain, and Jianfeng Gao.
\newblock Large language models: A survey, 2024.

\bibitem{hu2021loralowrankadaptationlarge}
Edward~J. Hu, Yelong Shen, Phillip Wallis, Zeyuan Allen-Zhu, Yuanzhi Li, Shean Wang, Lu~Wang, and Weizhu Chen.
\newblock Lora: Low-rank adaptation of large language models, 2021.

\bibitem{rafailov2023direct}
Rafael Rafailov, Archit Sharma, Eric Mitchell, Christopher~D Manning, Stefano Ermon, and Chelsea Finn.
\newblock Direct preference optimization: Your language model is secretly a reward model.
\newblock In {\em Thirty-seventh Conference on Neural Information Processing Systems}, 2023.

\bibitem{dettmers2023qloraefficientfinetuningquantized}
Tim Dettmers, Artidoro Pagnoni, Ari Holtzman, and Luke Zettlemoyer.
\newblock Qlora: Efficient finetuning of quantized llms, 2023.

\end{thebibliography}

\end{document}